# Earth Similarity Index and Habitability Studies of Exoplanets

*A Dissertation Submitted in Partial Fulfillment of the
Requirements for the Award of the Degree of*

Master of Philosophy
in
Physics

By
Madhu Kashyap J
(Reg No 1540016)

Under the Guidance of
Shivappa B Gudennavar
Associate Professor

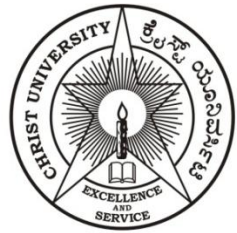

CHRIST UNIVERSITY
BENGALURU, INDIA

Declared as Deemed to be University under Section 3 of UGC Act 1956

Department of Physics and Electronics
CHRIST UNIVERSITY
BENGALURU, INDIA

**February 2017**

# Approval of Dissertation

Dissertation entitled "**Earth Similarity Index and Habitability studies of Exoplanets**" by Madhu Kashyap Jagadeesh (Reg No 1540016) is approved for the award of the degree of Master of Philosophy in Physics.

Examiners:

1. \_\_\_\_\_\_\_\_\_\_\_\_\_\_\_\_\_    \_\_\_\_\_\_\_\_\_\_\_\_\_\_\_\_\_

2. \_\_\_\_\_\_\_\_\_\_\_\_\_\_\_\_\_    \_\_\_\_\_\_\_\_\_\_\_\_\_\_\_\_\_

Supervisor:

\_\_\_\_\_\_\_\_\_\_\_\_\_\_\_\_\_    \_\_\_\_\_\_\_\_\_\_\_\_\_\_\_\_\_

Chairman:

\_\_\_\_\_\_\_\_\_\_\_\_\_\_\_\_\_    \_\_\_\_\_\_\_\_\_\_\_\_\_\_\_\_\_

Date: \_\_\_\_\_\_\_\_\_\_

Place: Bengaluru

(Seal)

# Declaration

I, Madhu Kashyap Jagadeesh, hereby declare that the dissertation titled "**Earth Similarity Index and Habitability studies of Exoplanets**" is a record of original research work undertaken by me for the award of the degree of Master of Philosophy in Physics. I have completed this study under the supervision of Dr Shivappa B Gudennavar, Associate Professor, Department of Physics and Electronics, Christ University, Bengaluru.

I also declare that this dissertation has not been submitted for the award of any degree, diploma, associateship, fellowship or other title. It has not been sent for any publication or presentation purpose. I hereby confirm the originality of the work and that there is no plagiarism in any part of the dissertation.

Place: Bengaluru
Date:

Madhu Kashyap J.
Reg No 1540016
Department of Physics and Electronics
Christ University, Bengaluru-560029.

# Certificate

This is to certify that the dissertation submitted by Madhu Kashyap Jagadeesh (Reg No 1540016) titled "**Earth Similarity Index and Habitability studies of Exoplanets**" is a record of research work done by him during the academic year 2015-2016 under my supervision in partial fulfilment for the award of Master of Philosophy in Physics.

This dissertation has not been submitted for the award of any degree, diploma, associateship, fellowship or other title. It has not been sent for any publication or presentation purpose. I hereby confirm the originality of the work and that there is no plagiarism in any part of the dissertation.

Place: Bengaluru
Date:                                   Dr Shivappa B Gudennavar
                                        Associate Professor of Physics and Electronics,
                                        Associate Director, Centre for Research
                                        Christ University, Benagluru-560029

Dr. George Thomas C
Professor & Head,
Department of Physics and Electronics
Christ University, Bengaluru-560029

# Acknowledgement

It is my real duty to thank my guru (Research Supervisor) Dr Shivappa B Gudennavar, who is the real G-force towards my work. He has taught me so many techniques not only to solve research problems, but also in the several aspects of admiring life itself.

Also, I would be honoured to thank Dr Bubbly S G, MPhil co-ordinator, for her constant support. Dr. Margarita S from Indian Institute of Astrophysics (IIA) has played an integral part during my research work, so a special Russian way of thanks to her "blagodaryu".

A special thanks to the HOD of Physics Dr George Thomas C for guiding me in his own humble way, also want to thank all the faculty members of Christ University (namely: Dr K T Paul, Dr. B. Manoj, Dr A G Kunjomana, Dr S Ravichandran).

I want to thank so many people, who are the real back bone for my research as well as pushed me in the right direction to become a deserving research scholar. Primarily, I want to thank my mother Mrs. Sathyavathi whole heartedly, for the way she has raised me towards the beacon of light. Both the parents play a major role and I don't hold back to say that my father Mr. Jagadeesh is my hero. He has given me everything I need than any other person in this whole multi-verse. He is my inspiration and it is his dream to see me as a wise person (Dr.) and this MPhil degree is the penultimate step in this ladder of fulfilment. I would like to thank the loved ones who made this successful Vani and Karthik.

It's an honour and privilege to thank Dr. Sr. Elizabeth principal of JNC for her blessing towards my work. My Physics department at JNC has given me 3 special mothers in the form of Ms.Sreedevi, Mrs. Thankamma, and Mrs. Carol for actual spirit uplifting wisdom. People who have kept my mind in an enlightened state during my real work my brother Karthik, Prashanth, Kiran, Ajin, Nithin and Arjun. Time to thank the heroes from behind the scene, my 9th grade teacher Nagendra sir, Madhusudan sir, Dr. B.S. Shylaja, APJ sir, Dr. Sudha Gopalakrishna mam, Dr. Sovan Ghosh.

Family plays a major role in moulding a person, heartfelt thanks to all the members: paternal uncle Manjunath, my meternal uncles Nagraj, Venkatesh Murthy, all my cousin brothers and sisters (namely: Shreyas, Ullas, Varsha, Suhas, Shardha, Nagashree, Rekha, Manjula, etc.). A tearful demise mentioning's at feet to my paternal grandfather Bhyrappa, uncle Nagraj, grandmother Seethalakshmi, maternal grandmother Rajamma etc.

I would like to thank Professor Jayant Murthy of Indian Institute of Astrophysics (IIA), Bengaluru for the fruitful discussions.

This research has made use data from Extrasolar Planets Encyclopaedia, Exoplanets Data Explorer the Habitable Zone Gallery and the NASA Exoplanet Archive.

# Abstract

Study of exoplanets has been of considerable interest for Astronomers, Planetary Scientists and Astrobiologists to look for alien life. Analysis of huge planetary data from space missions such as CoRoT and Kepler is directed ultimately at finding a planet similar to Earth- the Earth's twin, and looking for potential habitability. The Earth Similarity Index (ESI) is defined to find the similarity with Earth, which ranges from 1 (Earth) to 0 (totally dissimilar to Earth). ESI can be computed using four physical parameters of a planet, namely radius, density, escape velocity and surface temperature. The ESI is further sub-divided into interior ESI (geometrical mean of radius and density) and surface ESI (geometrical mean of escape velocity and surface temperature). The challenge here is to determine which parameter of exoplanet is important in finding this similarity; how exactly the individual parameters entering the interior ESI and surface ESI are contributing to the global ESI. The surface temperature entering surface ESI is a non-observable quantity and what we know is only equilibrium temperature of exoplanets. In this work, we have collated a comprehensive data on 3566 exoplanets from various sources. We have established a relation between surface and equilibrium temperatures using the data available for the solar system objects to address the difficulty in determining surface temperature. We then estimated surface temperature of all these exoplanets using this relation for further analysis of ESI. From the analysis, we have found 20 Earth-like exoplanets with ESI value above 0.8, which is set as the threshold, and these 20 exoplanets may be potentially habitable planets.

We are also interested in Mars-like planets to search for planets that may host the extreme life For example, methane on Mars may be a product of the methane-specific extremophile life form metabolism. We have proposed a new approach, called Mars Similarity Index (MSI), to identify planets that may be habitable to the extreme forms of life. MSI is defined in the range between 1 (present Mars) and 0 (dissimilar to present Mars) and uses the same physical parameters as that of ESI. Our study revealed that Moon with MSI of 0.75, Earth with 0.68 and the next closest exoplanet Kepler-186 f with 0.69 can be potentially habitable planets for extremophiles, which may further evolve at later times.

We introduced another new approach to study the potential habitability of exoplanets based on Cobb-Douglas Function, multi-parametric function. This did not yield any encouraging results.

# List of Publications

Similarity indexing of exoplanets in search for potential habitability: application to Mars-like worlds.
J M Kashyap, S B Gudennavar, Urmi Doshi and M Safonova
*Astrophysics and Space Science*, 2017 (Under Review)

# Contents

# List of Figures

# List of Tables

# Chapter 1

# Introduction

## 1.1 Introduction

Study of exoplanets and their habitability has been a very fascinating area of research in recent years. Presently many efforts have been devoted towards the discovery of the potential world for habitability via of space missions. The first spacecraft CoRoT mission was launched in 2006. The recent and the most popular launch is Kepler spacecraft by NASA (2009). The future missions (James Webb Space Telescope mission 2018 and PLATO 2024) are being planned to understand the climatic conditions of the exoplanets and to verify certain hypothesis like rocky planets may be habitable relaying on their air conditioning system (Carone, Keppens and Decin 2016).

There have been several methods adopted by astronomers to detect exoplanets; namely, Radial Velocity Technique, Transit and Occultation, Microlensing, Direct Imaging etc (Seager et al. 2007; Seager and Deming 2010; Anglada-Escud et al. 2013; Sengupta 2016). The details are presented in Section 2.1. The parameters such as mass, radius, temperature and escape velocity of exoplanets are obtained by employing any of these techniques. Having obtained these parameters, it is necessary to understand how similar these exoplanets to Earth to look for potential habitability. A multi-parameter indexing, say, Earth Similarity Index (ESI) has been defined as a number between zero (no similarity) and one (identical to Earth) to assess the Earth likeness for solar and exoplanets (Schulze-Makuch et al. 2011a; Mascaro 2011; Biswas et al. 2016). The ESI scale depends on the radius, density, escape velocity and surface temperature of exoplanets. The exoplanet having ESI value above 0.8 is considered to be Earth like and expected to have similar size, composition and atmosphere as that of Earth to support terrestrial life. Recent confirmation of Kepler 22b near a G-type star has set a threshold value of ESI as 0.9 along with Gliese 581g for the potentially habitable world.

Recently a better approach was described to understand the habitability of exoplanets in terms Planetary Habitability Index (PHI) with varying chemical composition (Schulze-Makuch et al. 2011a). PHI approach requires the presence of a stable substrate with appropriate chemistry to hold a liquid solvent, which supports life. But knowing the presence of a stable substrate having these features on all exoplanets is a challenging task.

But the efforts to propose new methods to explore habitable planets are intense. Saha et al. (2016)

have developed a novel revenue optimization model of a data center under sustainable budgetary constraints employing Cobb-Douglas production function. The function is widely used to represent the relationship between two or more inputs and the corresponding output parameters. This can potentially be used to study the habitability of exoplanets with parametric data available from space missions.

In the present investigation, we have made an attempt to collate the data of 3566 exoplanets available online as archives and analyzed it for Earth Similarity Index (ESI), Mars Similarity Index (MSI) and habitability of exoplanets using Cobb-Douglas function. We have studied how exactly the individual parameters entering the interior ESI (geometrical mean of radius and density) and surface ESI (geometrical mean of surface temperature and escape velocity), are contributing to the global ESI using graphical analysis. The mean surface temperature parameter entering into the surface ESI is non-observable quantity. In the PHL-EC data, maintained by the PHL, this parameter is obtained by adding a correction factor of 30 K based on the Earth's greenhouse effect. The main limitation of this method is that it is not consistent with all the given exoplanets. Because it does not justify the fact that a planet requires a rocky surface and an atmosphere to determine the surface temperature. In view of this, in this work, we have established a relation between surface and equilibrium temperatures using data available for the solar system objects, which have rocky surface and atmosphere. Using this relation, we estimated mean surface temperature of exoplanets for analyzing ESI. From our study, we found that 20 Earth-like exoplanets with ESI value above 0.8 are potentially habitable planets.

At present, though Earth is the only place where life exists, there is a good plausibility that life could have existed on ancient Mars, i.e. in the past (Abramov and Mojzsis 2016). Therefore, scientists are also interested in Mars-like planets to search for extremophile life forms, such as the ones living in extreme environments on Earth. For example, methane atmosphere on Mars is one of the requirements for the existence of a methane-specific extremophile life form, which was detected by the curiosity rover (Grotzinger et al. 2015). In view of these implications, we have introduced a new indexing parameter, the Mars Similarity Index (MSI) to study the extremophile life form in Mars-like conditions (Onofri et al. 2015; Hu et al. 2016). The MSI scale ranges between one (present Mars) to zero (no similarity to present Mars). The similar graphical analysis was carried out for the MSI to find the most contributing parameter to global MSI. From the study, it is clear that Moon (MSI= 0.75), Earth (MSI = 0.68) and Kepler-186f (MSI = 0.69) can be considered potentially habitable planets with MSI threshold value as 0.6.

We have also used Cobb-Douglas production function to compute the habitability of exoplanets making use of the physical parameters available in our data set. This did not yield any improved results as we observed inappropriate patterns.

The dissertation is divided into four chapters: Chapter 1 gives the introduction to the study of exoplanets, identifies current status, the gaps and the proposed work in brief. In Chapter 2, we describe briefly the techniques employed for detecting exoplanets and determining the physical parameters. In this chapter, we also present the literature review on the studies of exoplanets. Chapter 3 deals with the compilation of data on exoplanets from various sources and analysis procedure in terms of ESI, MSI and planetary habitability function. In Chapter 4, we present the results, discussion, conclusions and future work.

# Chapter 2

# Detection of Exoplanets and Literature Survey

## 2.1. Exoplanets and the Detection Techniques

Exoplanet is an external harbouring planet outside our solar system. The search for exoplanets and life on them was initiated a few decades ago. But the systematic scientific approach in this direction was begun in 1995 with the discovery of the first confirmed exoplanet orbiting around the 51 Pegasi. Since then many exoplanets have been discovered using one or the more methods at a time to confirm the detection (Mayor and Queloz 1995). We explain here a few methods which are popularly used by astronomers to detect exoplanets: Radial Velocity or Doppler Method, Transit and Occultation, Gravitational Microlensing, Direct Imaging etc (Seager et al. 2007; Jones 2008; Seager and Deming 2010; Sengupta 2016).

### 2.1.1. Radial Velocity or Doppler Technique

Doppler shift technique is one of the popularly used methods to detect exoplanets and to estimate their physical parameters (Sengupta 2016). When we analyze the spectrum of starlight, we see that it consists of several dark lines appearing due to the absorption of light by different chemical species of the atmosphere of the star. When two stars orbit each other, they rotate around a common center of mass (barycenter of the system) situated at the line joining the centers of the two stars. If the mass of one star is very small compared to the other, the barycenter goes inside the heavier star. The situation becomes same for the small planet orbiting around a massive star. As a result of rotation of planet and the host star around the barycenter of the system, the star wobbles. If the orbital plane is such that the star moves periodically towards and away from the observer giving rise to the Doppler shift in the spectrum of star (Fig. 2.1). When the spectrum is analyzed, we see red and blue shift of the absorption lines. The velocity of the star induced by the planet is called radial velocity of the star, which can be obtained by calculating the shift of a particular absorption line. The relationship between shift of the absorption line and the radial velocity is given by

$$\frac{\Delta \lambda}{\lambda} = \frac{v}{c} \qquad (2.1)$$

where $\Delta \lambda$ is the shift in the emitted wavelength, v is the radial velocity and c is the velocity of light. The radial velocity of a star is related with the project mass of the planet as.

$$v = \left(\frac{2\pi G}{P}\right)^{1/3} \frac{M_P \sin(i)}{M_S^{2/3}} \quad \text{for } M_P \ll M_S. \tag{2.2}$$

Where $M_P$ and $M_S$ are the mass of the planet and the host star respectively, P is the orbital period of the planet, i is the orbital inclination angle, G is the gravitational constant. By knowing the relation between the radial velocity of host star and the mass of the planet, one can determine the mass of the planet. The radial velocity confirms the presence of the planet and provides the projected mass of the planet. Nearly half of the exoplanets were discovered using this technique (Schneider 2012).

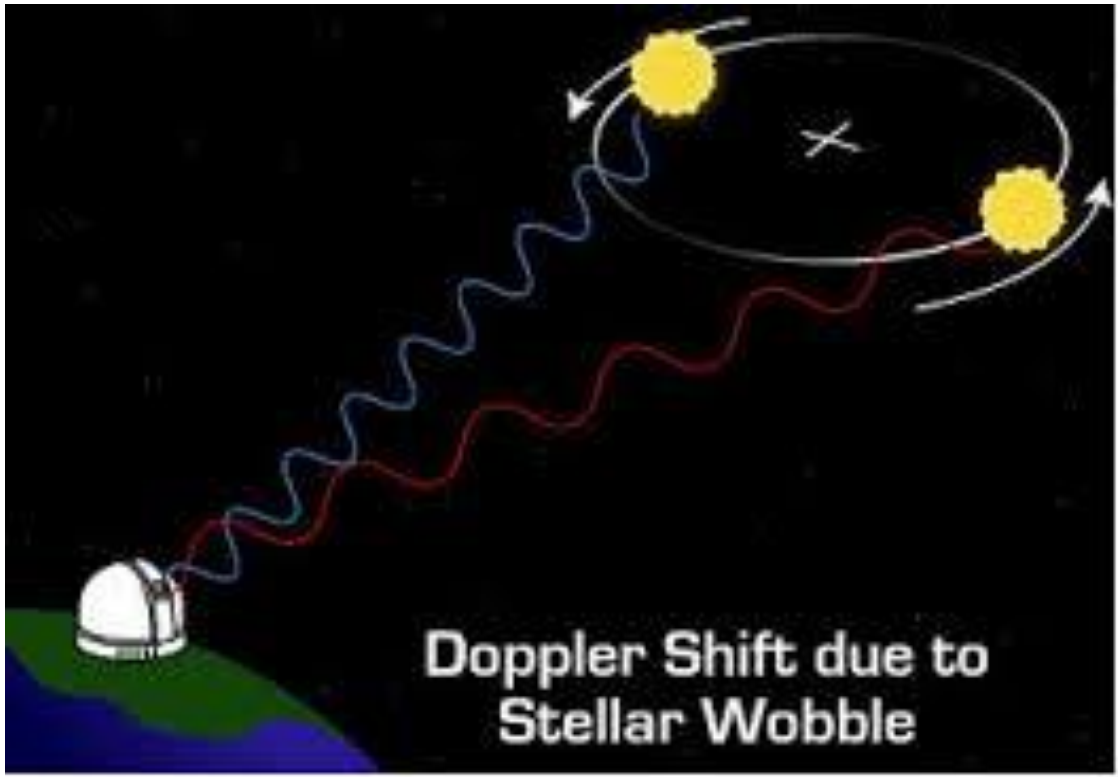

Figure 2.1: Radial Velocity Technique (Sengupta 2016)

### 2.1.2. Transit and Occultation Technique

When the planet passes between a star and an observer, the star's apparent brightness decreases. The variation in the apparent brightness of the star implies the presence of a planet (Fig. 2.2). The ratio of decrease in luminosity ($\Delta L$) to its actual luminosity is given as, $\Delta L / L = R_p^2 / R_s^2$. Here, $R_P$ is the radius of the planet and $R_S$ is the radius of the star. The radius of a planet can be determined accurately using this technique if size of the star is known. However, this technique is suitable when the orbit of planet is inclined to a large angle so that it can be viewed edge-on. Transit of a planet will work only if the orbital inclination angle $i$ is larger than $\cos^{-1}[(R_S + R_P)/a]$. Transit length is $2l = 2[(R_S + R)^2 - b^2]^{1/2}$, where $b = a \cos i$ and $a$ is the orbital separation between the star and the planet. The total transit duration is $T = P \sin^{-1}(l/a)/2\pi$, $P$ being the orbital period of the planet (Sengupta 2016). The Kepler and CoRoT space missions have sensitive instruments to detect the dip in the brightness with respect to time. Smaller planets will have smaller dip, while the larger planets have larger dip. The Kepler mission has discovered more than 1000 exoplanets using this method.

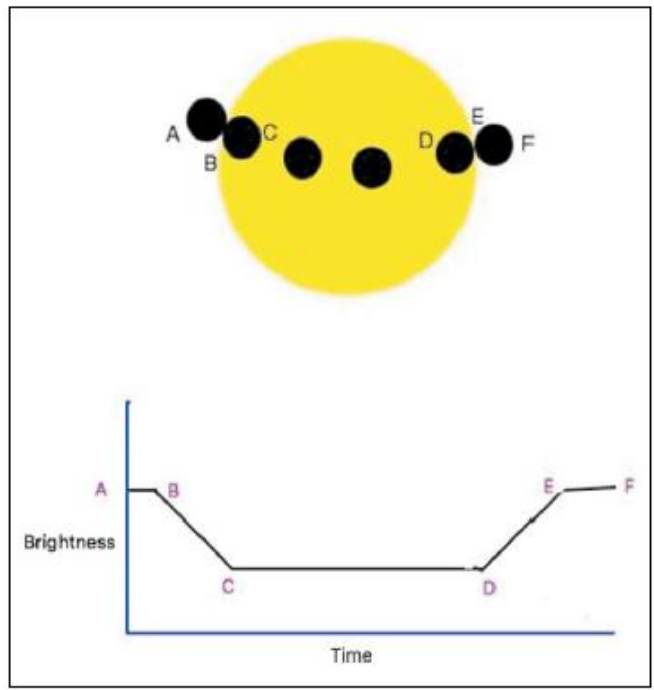

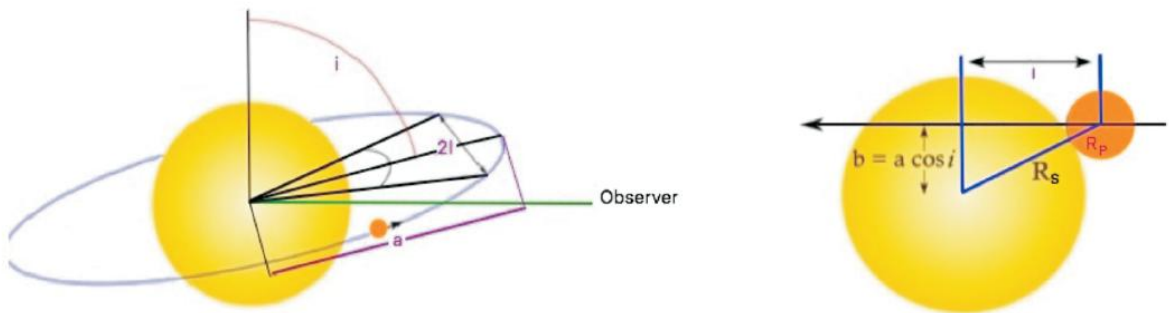

Figure 2.2: Transit and transit of geometry of exoplanets (Sengupta 2016).

## 2.1.3. Microlensing Technique

Albert Einstein proposed that the space-time surrounding a massive object is curved due to the strong gravitational attraction of the object. When light from distant star passes near a massive object, it bends. The gravity of the massive object acts like a lens, which results in sudden increase in the brightness. It the star has a planet, it acts like a lens. Since the process occurs for a period of 2-4 days, the observation cannot be made during the daytime. The event is observed by using many telescopes placed at different parts of the Earth. This method provides only the ratio of the mass between the planet and its parent star but not the individual parameters as that of other methods (Sengupta 2016).

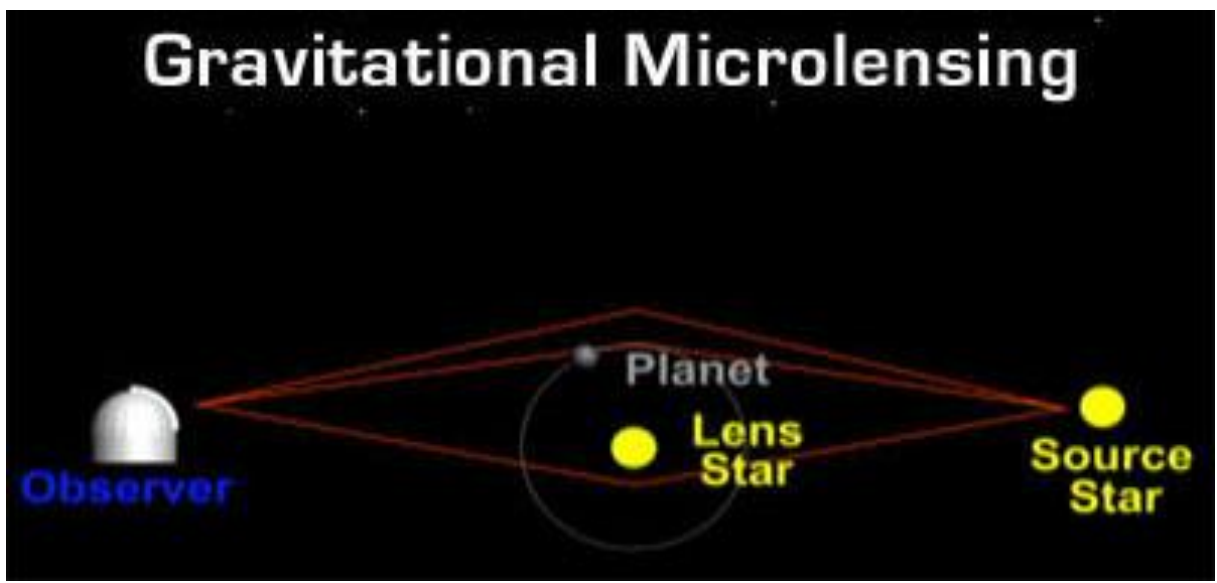

Figure 2.3: Gravitational Microlensing (Sengupta 2016)

## 2.1.4. Direct Imaging Technique

This technique is based on direct observation of light from planets. But it has difficulty due to the intense glare of the star when the distance between planet and host star is very close. Recently a few exoplanets have been imaged directly, which are far away from their host stars (distance of 10–30 AU or more) blocking the star light using a Coronagraph.

Having got introduced to the different methods which are in use to detect exoplantes, we will present the chronological progress of studies on exponents in the Section 2.2 below.

## 2.2. Literature Survey

Previously, similarity functions are applied in various other systems like ecology by Bray and Curtis (1957). Estimation of affinities in prairie in ecology was done by Looman and Campbell (1960). For the mathematical formulation of the above said similarity indices, the community study pattern was introduced from Bloom (1981). The empirical formula to understand the huge statistical distributions was discovered by Douglas called as Cobb-Douglas function in 1976. In 1989, Melosh and Vickery studied the impact erosion of the primordial atmosphere of Mars.

The efforts to look for extraterrestrial intelligence were initiated decades ago, the real breakthrough started with the first discovery of an exoplanet in the constellation of Pegasus in 1992 by Aleksander Wolszczan (Wolszczan and Frail 1992; Wolszczan 1994; Wolszczan 1997). The host star is 51 Pegasi and the exoplanet is called 51 Pegasi b (Mayor and Queloz 1995). The probability of complex life in circumstellar habitable zone outside the earth was proposed by Kasting (1993). Gaidos et al. (2005) studied the mass definitional limit of exoplanets and proposed the right range for exoplanets to lie between 0.1 to 10 times mass of Earth for sustaining atmosphere for life. Tung, Bramall and Price showed the microbial origin of excess methane in glacial ice and thus implying life on Mars (2005).Tree diversity analysis, through manual and software for statistical methods in ecological and biodiversity studies by Kindt and Coe (2005). The past and future planet hunt from astrometry by Alexander was done in 2006. The distance scale for similarities has been compiled in the form of a dictionary by Deza and Deza (2006). Sotin, Grasset and Mocquet *(*2007) proposed a relationship between mass and radii of exoplanets by studying the internal structure of Earth-like planets and large icy satellites of the solar system. The similarity function is developed further by comparison of probability functions Cha (2007). The day-night variation of HD 189733b extrasolar planet has been recorded in the form of a map by Knutson et al. (2007). Kounaves (2007) proposed a way to Mars like planet, which would be habitable for extremophiles. Seager et al. (2007) shredded the light towards mass range for rocky exoplanets, radius range for rocky exoplanets and mass-radius relationships for solid exoplanets. Nesvorný and Morbidelli (2008) proposed a method for the determination of mass and orbit of exoplanets from transit timing variations of exoplanets. The search methods, discoveries, and prospects for astrobiology has been well described by Jones (2008). Baraffe, Chabrier and Barman (2010) reported the determination of physical properties like mass, density, temperature, escape velocity, radius and period of extra-solar planets. Further, the two tier system of ESI and PHI was studied by Schulze-Makuch et al. (2011a), which is useful for understanding the similarity of exoplanets to earth in many respects. ESI is a logical breakpoint in the nomenclature of exoplanets (Mascaro 2011). Waltham et al. (2011) described the anthropic selection and habitability of planets orbiting M and K dwarf stars. In 2012, Fressin et al. discovered two earth sized planet near the vicinity of Kepler-20 system.

The hunting for exoplanets has been continued and the Planetary Habitability Laboratory (2012) has been listing the details of more than 3000 exoplanets discovered by now. Identifying the false positive detections of exoplanets in transit method is a difficult task. Since astronomers know only a small fraction of the Kepler and CoRoT candidates that can be established as genuine planets, the only way to determine the false positive probability (FPP) of these candidates is by developing new models and theories (Santerne et al. 2013). The ecological application for similarity functions is analyzed in multivariate way (Greenacre and Primicerio 2013).

Further, an attempt to understand the recycling mechanisms of atmospheric $CO_2$ and $CH_4$ in the potential water-rich planets was made by Kaltenegger et al. (2013). A multi-planetary system was found in the habitable zone of GJ 667C (Anglada-Escud et al. 2013). The atmosphere scales of Mars gives the details regarding the temperature variation on Mars (Barlow 2014). Hadden and Lithwick (2014) have measured densities and eccentricities of Kepler-139 system from transit time varying method. The ancient Mars analysis has been done by Bell et al. (2015). Deposition, exhumation and paleoclimate of ancient lake deposit on Mars have been studied by Grotzinger et al. (2015). Methane detection and variability at Gale crater has be analyzed by Webster et al. (2015).

Subsequently, Aronson, Walden and Piskunov (2015) developed an atmospheric model to determine the water rich like planets. Onofri et al. (2015) showed that the antartic cryptoendolithic fungi could survive in stimulated Martian conditions. The thermal implications on Mars have been well studied by Abramov and Mojzsis (2016). The climate and tidal variation has been modeled and studied by Carone, Kappens and Decin (2016). Methane presence on the surface of Mars has been well hypothesized b Hu et al. (2016). The climate of ancient Mars plays a major role in studying its habitability (Wordsworth, 2016). Orbital evidence for more widespread carbonate-bearing rocks on Mars has been studied by Wray et al. (2016). The relook of ESI and habitability zone search has been extended by Biswas et al. (2016).

From the literature, it is clear that study on exoplanets, with the progress in new technologies, is drawing considerable attention of the astrophysicists and astrobiologists in the recent years in many respects. In the present work, we planned to accomplish the following objectives,

- To compile the data on exoplanets available online as archives
- To study Earth Similarity Index (ESI),
- To study the Mars Similarity Index (MSI) and
- Habitability of exoplanets using Cobb-Douglas function.

# Chapter 3
# Compilation of Data and Analysis

## 3.1. Data

The Planetary Habitability Laboratory, University of Pierto Rico at Arecibo has been doing a commendable job of maintaining data on observed and modeled parameters for all currently confirmed as well as unconfirmed exoplanets (PHL's Exoplanets Catalog). This catalog consists of stellar and planetary parameters such as habitability assessments and planetary classifications.

Though PHL's Exoplanets Catalog contains 3635 confirmed exoplanets (as of January 2017), some of them do not have all the required input parameters for calculating global ESI and MSI. In the present work, we supplemented the missing data by compiling after the search through the catalogs: Habitable Zone Gallery, Open Exoplanet Catalogue, Extrasolar Planets Encyclopedia, Exoplanets Data Explorer and NASA Exoplanet Archive. In addition, we have discarded the entries with unrealistic values. For example, some of the planets had the equilibrium/surface temperatures of less than 3.2 K, and some had the density values of around 500 EU. In such cases, we have done extensive search through all available exoplanet catalogs and literature on exoplanets study, and supplemented those missing values. As an example, there is a lot of confusion with the available data on Kepler-53c, Kepler-57c and Kepler-59b planets. Their densities are listed in the PHL-EC as 162, 573.18 and 492 EU, respectively. In such cases, the density of, say, Kepler 57c, becomes 21 times the density of the Sun's core. There is obviously a mistake in the retrieved data. We have searched through the catalogs and found that for Kepler 53c, the mass of 5007.56 EU used in calculating that density was, in fact, an upper limit from the stability analysis (Steffen et al. 2013). It was subsequently updated to $35.4^{+19}_{-14.8}$ EU with nearly the same value for the radius (Hadden and Lithwick 2014). Using this number, the density becomes 1.169 EU. Similarly, for Kepler 57c, the density of 573.18 EU was obtained using the upper limit on mass of 2208.83 EU which, after updating the mass to $7.4^{+9.4}_{-6.3}$ EU with essentially the same radius, fell within the normal range, 1.139 EU (Hadden and Lithwick 2014). Accordingly, we have corrected the data for these planets in our compiled catalog. Some of the entries had to be removed owing to the absence of available data (or very conflicting values), which left us with 3566 exoplanets for our analysis. A sample data is presented in Table 3.1 and the complete data set is made available online for benefit of scientific community (http://dx.doi.org/10.17632/c37bvvxp3z.6).

Table 3.1: Sample set of data

| Names | Radius (EU) | Density (EU) | Surface Temperature (K) | Escape Velocity (EU) |
|---|---|---|---|---|
| Kepler-438 b | 1.12 | 0.90 | 312 | 1.06 |
| GJ 667Cc | 1.54 | 1.05 | 280 | 1.57 |
| Kepler-296 e | 1.48 | 1.03 | 303 | 1.50 |
| Kepler-442 b | 1.34 | 0.97 | 265 | 1.32 |
| 1RXS 1609 b | 19.04 | 0.64 | 11.4 | 15.2 |
| Kepler-62 e | 1.61 | 1.080 | 296 | 1.68 |
| GJ 832 c | 1.69 | 1.120 | 287 | 1.79 |
| Kepler-452 b | 1.63 | 1.090 | 296 | 1.70 |
| K2-3 d | 1.52 | 1.040 | 319 | 1.55 |
| GJ 667C f | 1.40 | 0.990 | 252 | 1.39 |
| Kepler-283 c | 1.81 | 1.180 | 282 | 1.97 |
| tau Cet e | 1.59 | 1.070 | 319 | 1.64 |
| KIC-5522786 b | 1.23 | 0.940 | 344 | 1.19 |
| GJ 180 c | 1.77 | 1.160 | 272 | 1.90 |
| HD 85512 b | 1.50 | 1.030 | 336 | 1.53 |
| GJ 180 b | 1.89 | 1.220 | 303 | 2.09 |
| Kepler-440 b | 1.86 | 1.200 | 308 | 2.04 |
| GJ 682 b | 1.60 | 1.080 | 332 | 1.66 |
| GJ 163 c | 1.83 | 1.190 | 313 | 1.99 |

## 3.2 Data Analysis

We have compiled the data on various aspects of 3566 exoplanets. We shall analyze these data for the Earth Similarity Index (ESI) and Mars Similarity Index (MSI) to understand the potential habitability of those exoplanets. The details of the analysis procedure are presented in the following paragraphs.

### 3.2.1 Mathematical formulation for ESI and MSI

Distance/similarity measurements are widely used in classification of objects in various disciplines (Deza and Deza 2006). Here the distance 'd' is represented as dissimilarity and proximity is equivalent to similarity 's'. Mathematically, the concept of distance is a metric one- a measure of a true distance in Euclidean space $R^n$ This problem is usually addressed by using Minkowski's space of $L_p$ form (Cha 2007), in which p-norm stands for finite n-dimensional vector space,

$$d = (\sum_{i=1}^{n}|p_i - q_i|^p)^{1/p} \qquad (3.1)$$

where the Manhattan, $L_1$ distance,

$$d_{Man} = \sum_{i=1}^{n}|p_i - q_i| \qquad (3.2)$$

and Euclidean $L_2$ distance,

$$d_{Euc} = (\sum_{i=1}^{n}(p_i - q_i))^{1/2} \qquad (3.3)$$

are the special cases. Here $p_i$ and $q_i$ are the coordinates of p and q with dimension i and i = 1, 2, 3…., n. $L_1$ form has an advantage that it can be decomposed into contributions made by each variable being the sum of absolute differences. For example, for the $L_2$ form, it would be the decomposition of the squared distance. Our interest is to find similarities between different planets based on their various characteristics. In such cases, the Bray-Curtis distance is the most widely used scale (Bray and Curtis 1957; Greenacre and Primicerio 2013). Bray-Curtis is a modified Manhattan distance, where the summed differences between the variables are standardized by the summed variables of the objects,

$$d_{BC} = \frac{\sum_{i=1}^{n}|p_i-q_i|}{\sum_{i=1}^{n}(p_i+q_i)} \tag{3.4}$$

Here $p_i$ and $q_i$ are two different precisely measurable quantities between which the distance is to be measured, and n is the total number of variables. The Bray-Curtis scale assumes that the samples are taken from same physical measure, say, mass, or volume. It is because the distance is found from the raw counts, so that if there is a higher abundance in one sample comparing to the other, it is a part of the difference between the two samples. In Bray-Curtis scale, the interpretation is such that zero means the samples are exactly the same and one means they are completely disjoint. It shall be kept in mind that Bray-Curtis distance is not the true metric distance, it is a semi-metric distance (in which the distance between two distinct points can be zero), which is usually called dissimilarity, or ecological distance (Kindt and Coe 2005). The advantages in using the ecological distance is that differences between datasets can be expressed by a single statistic.

The intersection between two distributions is more widely used form of similarity (Looman and Campbell 1960). Most similarity measures for intersection can be transformed from the distance measure by the transformation technique, but not exclusively (Bloom 1981),

$$s_{BC} = 1 - d_{BC} = 1 - \frac{\sum_{i=1}^{n}|p_i-q_i|}{\sum_{i=1}^{n}(p_i+q_i)} \tag{3.5}$$

Here, the value of 0 means complete absence of relationships, and the value of 1 shows a complete matching of the two data records in the n-dimensional space (Schulz 2007).

Distances/similarities based on heterogeneous data can be found after a process of standardization-balancing of the contribution of different types of variables in an equitable way (Greenacre and Primicerio 2013). This is done by calculating the similarity for each set of homogeneous variables and then combining them using various methods, which will be described in the following paragraphs while showing the ESI calculation. Higher values in one set may influence the result of the Bray-Curtis similarity more dominantly and imply that these variables are more likely to discriminate between sets. Therefore, user-defined weighting is a convenient (though subjective) method for down-weighing the differences for a set of variables. In the present work, our aim is to compare sets of different variables of one planet with that of the reference value, for example, Earth, to find planets that are similar to Earth. Rewriting Eq. (3.5),

$$S = \left[1 - \frac{|x-x_0|}{(x+x_0)}\right]^{w_x} \tag{3.6}$$

where x is the physical property of the exoplanet, $x_0$ is the reference value, $w_x$ is the weight for this property, and the dimension n = 1, since we are constructing the index separately for each physical property. We find the weights by defining the threshold value (V) in the similarity scale for each quantity,

$$V = \left[1 - \frac{|x-x_0|}{(x+x_0)}\right]^{w_x} \tag{3.7}$$

In the literature (Bloom 1981), the similarity index ranging from zero to one is subdivided into 0.2 equal intervals, defining very low, low, moderate, high and very high similarity regions. The threshold is defined on this grounds, for example considering only very high similarity region with threshold value equal to V = 0.8. Fixing the threshold value V and defining the physical limits $x_a$ and $x_b$ as the permissible variation of a variable with respect to $x_0$ (ie, $x_a < x_0 < x_b$), we can calculate the weight exponents for the lower $w_a$ and upper $w_b$ limits,

$$w_a = \frac{\ln V}{\ln\left\{1-\left|\frac{x_0-x_a}{x_0+x_a}\right|\right\}} \quad \text{and} \quad w_b = \frac{\ln V}{\ln\left\{1-\left|\frac{x_b-x_0}{x_b+x_0}\right|\right\}} \tag{3.8}$$

The average weight is obtained by the geometric mean of lower $w_a$ and upper $w_b$ limits,

$$w_x = \sqrt{w_a \times w_b} \tag{3.9}$$

For the present studies, the Earth and Mars similarity indices are defined as,

$$ESI_x = \left\{1 - \left|\frac{x-x_0}{x+x_0}\right|\right\}^{w_x} \tag{3.10}$$

$$MSI_x = \left\{1 - \left|\frac{x-x_0}{x+x_0}\right|\right\}^{w_x} \tag{3.11}$$

where x is the physical parameter of the exoplanet such as radius or density and $x_0$ is the reference to Earth for ESI and to Mars for MSI.

The mean radius (R), bulk density (ρ), escape velocity (V$_e$) and surface temperature (T$_s$) of exoplanets are used as input parameters for computing similarity indices. These parameters, except the surface temperature (which left is in kelvin), are used in Earth Units (EU) for the calculation of ESI and in the Mars Units (MU) for the calculation of the MSI. The corresponding weight exponents for both ESI and MSI scales were computed adopting the threshold value V = 0:8, indicating very high similarity region. The weight exponents for the upper and lower limit of parameters were calculated for the Earth-like parameters (Schulze-Makuch et al. 2011a): radius range from 0.5 to 1.9 EU, mass range from 0.1 to 10 EU, density range from 0.7 to 1.5 EU, surface temperature range from 273 to 323 K, and escape velocity range from 0.4 to 1.4 EU, using Eqs. (3.8) and (3.9). Similarly, the weight exponents for the lower and upper limit of parameters are defined for the Mars-like conditions: radius range from 0.72 to 1.88 MU, mass range 0.514 to 9.30 MU, density range 0.89 to 1.402 MU, surface temperature range 233 to 418 K, and escape velocity range 0.85 to 2.23 MU. Where radius is 3390 km, density is 3.93 g/cm3, the mean surface temperature 240 K (Barlow 2014) and escape velocity is 5.03 km/s. The reasons behind the limits definitions are to have a rocky planet with lower limit in comparison to Mars (mass and radius are chosen as for Mercury, the smallest planet in our Solar System), and with Earth as the upper limit. The temperature range is chosen on the basis of the temperature known to be suitable for extremophile life forms, − 40 to +145 °C (Tung, Bramall and Price 2007). The corresponding weight exponents for MSI computations were calculated using the same method as for the ESI. The weight exponents calculated for ESI and MSI are presented in Table 3.2. The surface temperature required for the estimation of global ESI and MSI is determined as detailed in the following subsection.

Table 3.2: The parameters for ESI and MSI scale

| Planetary property | Reference values for ESI | Weight exponents for ESI | Reference values for MSI | Weight exponents for MSI |
|---|---|---|---|---|
| Mean radius (R) | 1 EU | 0.57 | 1 MU | 0.86 |
| Bulk density (ρ) | 1 EU | 1.07 | 1 MU | 2.10 |
| Escape velocity (V$_e$) | 1 EU | 0.70 | 1 MU | 1.09 |
| Surface temperature (T$_s$) | 1 288 K | 5.58 | 240 K | 3.23 |

### 3.2.1.1 Surface temperature estimation

Usually, the temperature of extrasolar planets is estimated from the calculated temperature of the star and other observational data such as distance to the star (Fressin et al. 2012). The equilibrium temperature of an exoplanet is determined using the intensity of the light the planet receives from its host star (Knutson et al. 2007). However, albedo entering the equilibrium temperature equation is generally not known and has to be assumed. Currently, the surface temperature required for the calculation of global ESI of rocky planets with atmospheres estimated using a correction factor of 30-33 K, based on the Earth's green-house effect (Schulze-Makuch et al. 2011a; Volokin and ReLlez 2016). The problem here is that the method is not consistent for all the exoplanets (i.e., it does not justify the fact that a planet requires a rocky surface and an atmosphere to calculate the surface temperature). In the present work, we have established a relation between mean surface temperature ($T_s$) and equilibrium temperature ($T_e$) using the data available for solar system objects, which have rocky surface and atmosphere, and estimated mean surface temperature ($T_s$) of exoplanets (Fig. 3.1). The relation thus established is given by,

$$T_s = 9.650 + 1.096 \times T_e \tag{3.12}$$

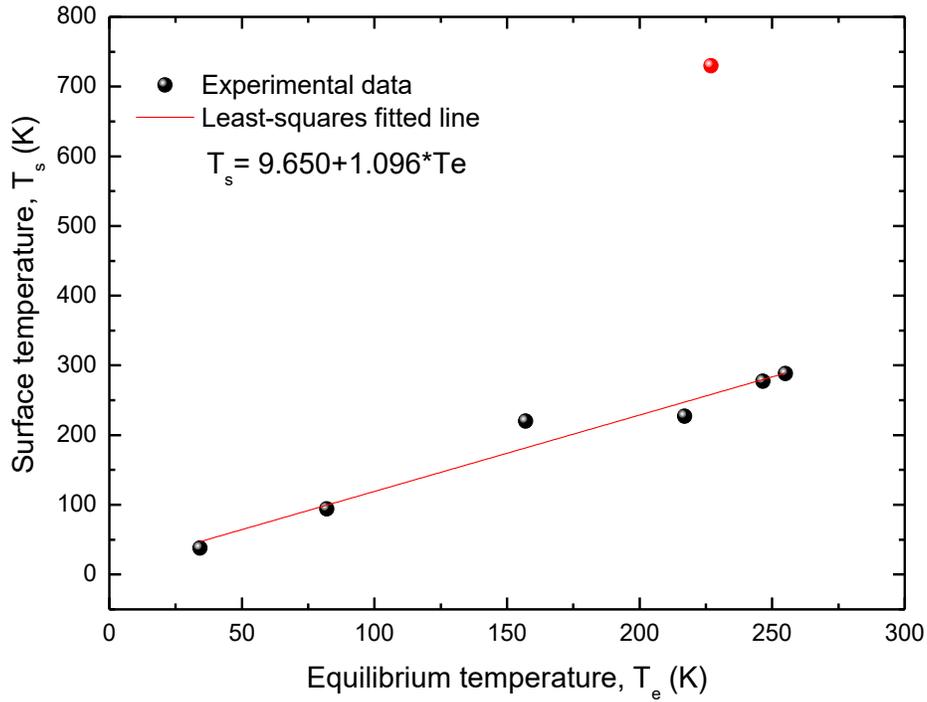

Figure 3.1: Equilibrium temperature versus surface temperature

Venus (dot due) due to its very high surface temperature does not obey the linear relationship. The equilibrium and surface temperature of a few Solar System objects along with the sample of our results for a few potentially habitable exoplanets are presented in Table 3.3. Since surface temperature of moon varies from 120 K on far side to 396 K under Sun, the surface temperature chosen for moon is 197 K (Volokin and ReLlez 2016) in Table 3.3.

### 3.2.1.2 ESI of exoplanets

In order to determine ESI of exoplanets, we converted all the input parameters to Earth Units (EU), except the surface temperature, which is expressed in Kelvin. The planets for which surface temperature information is not available; we have calculated the surface temperature using Eq. (3.12). The corresponding ESI for each parameter ($ESI_R$, $ESI_\rho$, $ESI_{Ts}$ and $ESI_{Ve}$) of the planet was calculated using Eq. (3.10). Using these ESI values, the interior ESI and surface ESI are determined using the following relations,

$$ESI_I = \sqrt{ESI_R \times ESI_\rho} \qquad (3.13)$$

$$ESI_S = \sqrt{ESI_{Ts} \times ESI_{ve}} \qquad (3.14)$$

Where, $ESI_R$, $ESI_\rho$, $ESI_{Ts}$ and $ESI_{Ve}$ are ESI values calculated for radius, density, surface temperature and escape velocity respectively. Then the global ESI is calculated using the relation,

$$ESI = \sqrt{ESI_I \times ESI_S} \qquad (3.15)$$

Table 3.3: Equilibrium and surface temperature

| Planet | Equilibrium temperature, $T_e$ (K) | Surface temperature, $T_s$ (K) | $T_s = 9.650 + 1.096 T_e$ (K) |
|---|---|---|---|
| Earth | 255 | 288 | 289.28 |
| Mars | 217 | 240 | 247.61 |
| Moon | 157 | 197 | 181.81 |
| Venus | 227 | 730 | 258.57 |
| Titan | 82 | 94 | 99.57 |
| Triton | 34.2 | 38 | 47.15 |
| GJ 667Cc | 246.5 | 277.4 | 279.96 |
| Kepler-442 b | 233 | Not known | 265.16 |
| Kepler-438 b | 276 | Not known | 312.31 |
| GJ 667 C f | 220.7 | Not known | 251.67 |

The sample calculation of ESI for Mars using the weight exponents in Table 3.2 and with the input parameters is shown below.

$R = 0.53 \times 6371 km = 3376.63 km$
$\rho = 0.71 \times 5.51 g/cm^3 = 3.91 g/cm^3$
$v_e = 0.45 \times 11.19 km/s = 5.03 km/s$
$T = 240 K$

The corresponding ESI parameters are

$$ESI_R = (1-|3376.63km - 6371km|/|3376.63km + 6371km|)^{0.57} = 0.81$$

$$ESI_\rho = (1-|3.91g/cm^3 - 5.51g/cm^3|/|3.91g/cm^3 + 5.51g/cm^3|)^{1.07} = 0.82$$

$$ESI_{v_e} = (1-|5.03km/s - 11.19km/s|/|5.03km/s + 11.19km/s|)^{0.7} = 0.72$$

$$ESI_{T_s} = (1-|240K - 288K|/|240K + 288K|)^{5.58} = 0.59$$

Interior ESI is $ESI_I = \sqrt{(0.81 \times 0.82)} = 0.82$

Surface ESI is $ESI_S = \sqrt{(0.72 \times 0.59)} = 0.65$

Global ESI is $ESI = \sqrt{(0.82 \times 0.65)} = 0.73$

The results of ESI calculation for all 3566 currently confirmed exoplanets are made available online (Kashyap, Safonova and Gudennavar 2017). The sample of results for a few planets is presented in Table 3.4.

Table 3.4: A sample of results of calculated ESI

| Names | Radius (EU) | Density (EU) | Surface temperature (K) | Escape velocity (EU) | $ESI_I$ | $ESI_S$ | ESI |
|---|---|---|---|---|---|---|---|
| Earth | 1.00 | 1.00 | 288 | 1.00 | 1.00 | 1.00 | 1.00 |
| Mars | 0.53 | 0.73 | 240 | 0.45 | 0.82 | 0.65 | 0.73 |
| Kepler-438 b | 1.12 | 0.90 | 312 | 1.06 | 0.95 | 0.88 | 0.91 |
| GJ 667Cc | 1.54 | 1.05 | 280 | 1.57 | 0.92 | 0.88 | 0.90 |
| Kepler-296 e | 1.48 | 1.03 | 303 | 1.50 | 0.93 | 0.86 | 0.89 |
| Kepler-442 b | 1.34 | 0.97 | 265 | 1.32 | 0.94 | 0.84 | 0.89 |
| Kepler-62 e | 1.61 | 1.080 | 296 | 1.68 | 0.90 | 0.87 | 0.88 |
| GJ 832 c | 1.69 | 1.120 | 287 | 1.79 | 0.89 | 0.88 | 0.88 |
| Kepler-452 b | 1.63 | 1.090 | 296 | 1.70 | 0.90 | 0.86 | 0.88 |
| K2-3 d | 1.52 | 1.040 | 319 | 1.55 | 0.92 | 0.79 | 0.85 |
| GJ 667C f | 1.40 | 0.990 | 252 | 1.39 | 0.94 | 0.77 | 0.85 |
| Kepler-283 c | 1.81 | 1.180 | 282 | 1.97 | 0.86 | 0.84 | 0.85 |
| tau Cet e | 1.59 | 1.070 | 319 | 1.64 | 0.91 | 0.78 | 0.84 |
| KIC-5522786b | 1.23 | 0.940 | 344 | 1.19 | 0.95 | 0.95 | 0.95 |
| GJ 180 c | 1.77 | 1.160 | 272 | 1.90 | 0.87 | 0.81 | 0.84 |
| HD 85512 b | 1.50 | 1.030 | 336 | 1.53 | 0.93 | 0.73 | 0.82 |
| GJ 180 b | 1.89 | 1.220 | 303 | 2.09 | 0.85 | 0.79 | 0.82 |
| Kepler-440 b | 1.86 | 1.200 | 308 | 2.04 | 0.85 | 0.78 | 0.82 |
| GJ 682 b | 1.60 | 1.080 | 332 | 1.66 | 0.90 | 0.73 | 0.81 |
| GJ 163 c | 1.83 | 1.190 | 313 | 1.99 | 0.86 | 0.77 | 0.81 |

Here, EU = Earth Units, where Earth's radius is 6371 km, density is 5.51 g/cm$^3$ and escape velocity is 11.19 km/s.

**3.2.1.3 MSI of exoplanets**

All the input parameters were converted to Mars Units (MU), except the surface temperature, before calculating MSI of exoplanets. The surface temperature was estimated using Eq. (3.12). Then MSI for each parameter (MSI$_R$, MSI$_\rho$, MSI$_{Ts}$ and MSI$_{Ve}$) of the planet was calculated using Eq. (3.10). Using these values, the interior MSI, surface MSI and global MSI are calculated using the relations,

$$MSI_I = \sqrt{MSI_R \times MSI_\rho} \qquad (3.16)$$

$$MSI_S = \sqrt{MSI_{Ts} \times MSI_{ve}} \qquad (3.17)$$

$$MSI = \sqrt{MSI_I \times MSI_S} \qquad (3.18)$$

Here MSI$_R$, MSI$_\rho$, MSI$_{Ts}$ and MSI$_{Ve}$ are MSI calculated for radius, density, surface temperature and escape velocity respectively. The sample calculation of MSI for Earth is shown below. The input parameters are

$R = 1.88 \times 3390 km = 6373 km$
$\rho = 1.40 \times 3.93 g/cm^3 = 5.5 g/cm^3$
$v_e = 2.23 \times 5.03 km/s = 11.2 km/s$
$T_s = 288 K$

The corresponding MSI parameters are

$MSI_R = (1-|6373km - 3390km|/|6373km + 3390km|)^{0.86} = 0.73$
$MSI_\rho = (1-|5.5g/cm^3 - 3.93g/cm^3|/|5.5g/cm^3 + 3.9g/cm^3|)^{2.10} = 0.68$
$MSI_{v_e} = (1-|11.2km/s - 5.03km/s|/|11.2km/s + 5.03km/s|)^{1.09} = 0.59$
$MSI_{T_s} = (1-|288K - 240K|/|288K + 240K|)^{3.23} = 0.73$

Interior MSI is $MSI_I = \sqrt{(0.73 \times 0.68)} = 0.70$

Surface MSI is $MSI_S = \sqrt{(0.73 \times 0.59)} = 0.66$

Global MSI is $MSI = \sqrt{(0.66 \times 0.70)} = 0.68$

The sample of results for a few planets is presented in Table 3.5, whereas the full data is available online (Kashyap, Safonova and Gudennavar 2017).

Table 3.5: A sample of results of calculated MSI

| Names | Radius (MU) | Density (MU) | Surface temperature (K) | Escape velocity (MU) | $MSI_I$ | $MSI_S$ | MSI |
|---|---|---|---|---|---|---|---|
| Earth | 1.88 | 1.40 | 288 | 2.23 | 0.70 | 0.66 | 0.68 |
| Mars | 1.00 | 1.00 | 240 | 1.00 | 1.00 | 1.00 | 1.00 |
| Moon | 0.51 | 0.85 | 197 | 0.48 | 0.77 | 0.66 | 0,71 |
| Kepler-186 f | 2.21 | 1.29 | 215 | 2.49 | 0.67 | 0.70 | 0.69 |
| Kepler-438 b | 2.11 | 1.27 | 312 | 2.36 | 0.72 | 0.60 | 0.66 |
| Kepler-442 b | 2.53 | 1.36 | 265 | 2.93 | 0.65 | 0.63 | 0.64 |

Here, MU = Mars Units, where radius is 3390 km, density is 3.93 g/cm$^3$, escape velocity is 5.03 km/s.

## 3.3 Planetary Habitability using Cobb Douglas Function

We have also made an effort to understand the habitability of exoplanets using Cobb–Douglas production function, which was proposed by Douglas (1976) to seek a functional form to relate estimates he had calculated for workers and capital (Saha et al. 2016).

Let the parameters for similarity index be R, $\rho$, $V_e$ and $T_s$, which represent radius, density, escape velocity and surface temperature respectively of a planet, with a, b, c, d being their corresponding input free parameters. The corresponding Cobb-Douglas function for calculating the potential habitability of an exoplanet is given by,

$$F = kR^a \rho^b v_e^c T^d, \text{ where, k is a constant.} \tag{3.19}$$

There can be different cases.

### (i) Increasing parametric similarity (IPS)

In this case, it is assumed that, $a + b + c + d > 1$ for which $a>0$, $b>0$, $c>0$ and $d>0$. This condition can be imposed on the function in Eq. (3.19) with the IPS input values from Table 3.6. Here, k = 1, $a = 0.29$, b = 0.20, c = 0.20 and d = 0.80 and the corresponding IPS value is obtained.

### (ii) Decreasing parametric similarity (DPS)

In this case, it is assumed that, $a + b + c + d < 1$ for which $a<0$, $b<0$, $c<0$ and $d<0$. Using this condition in Eq. (3.19) and the input values from Table 3.6, we can obtain k = 0.08, $a = 0.05$, b = 0.45, c = 0.04, d = 0.45 and the corresponding DPS values. Here k is chosen less than 1 in order to normalize the function between zero and one.

### (iii) Constant parametric similarity (CPS)

In this case, the values of *a*, b, c, and d is chosen such that $a + b + c + d = 1$ for which $a>0$, $b>0$, $c>0$ and $d>0$. Imposing this condition on the function in Eq. (3.19), the values k = 1, $a = 0.25$, b = 0.25, c = 0.25 and d = 0.25 are set and the corresponding CPS values are obtained.

The results (IPS, DPS and CPS) obtained for all the three cases are summarized in Table 3.6.

Table 3.6: A sample of results of calculated CDF

| Name | Radius (EU) | Density (EU) | Escape Velocity (EU) | Surface Temperature (K) | CDF (IPS) | CDF (DPS) | CDF (CPS) |
|---|---|---|---|---|---|---|---|
| Earth | 1.00 | 1.00 | 1.00 | 288 | 1.01 | 1.11 | 4.12 |
| Mars | 0.53 | 0.71 | 0.45 | 240 | 0.47 | 1.03 | 2.52 |
| KOI-1843b | 0.58 | 1.61 | 0.74 | 1536.6 | 0.70 | 1.10 | 5.74 |
| Kepler-444b | 0.4 | 0.74 | 0.35 | 972.4 | 0.36 | 1.04 | 3.17 |

The results, discussion, conclusions and the future work are presented in next chapter.

# Chapter 4

# Results, Discussion and Conclusions

## 4.1 Results and Discussion

We have refined and compiled the data on 3566 exoplanets taking the data from PHL's Exoplanets Catalog: PHL-EC. This compilation consists of the planetary data such as mean radius (R), bulk density (ρ), escape velocity ($V_e$) and surface temperature ($T_s$) (sample data in Table 3.1). The complete data set is made available online for benefit of scientific community. These parameters, except the surface temperature (which left is in kelvin), are used in Earth Units (EU) for the calculation of ESI and in the Mars Units (MU) for the calculation of the MSI. We have computed the similarity indices (ESI & MSI) using these values as input parameters as detailed in Chapter 4, which is made available online (which consists of data from Table 3.1, Table 3.4 and Table 3.5).

The results of ESI calculations are presented as a histogram of global ESI in Fig. 4.1. According to the PHL project, surface ESI is dominating the interior ESI, because the surface temperature weight exponent value is much higher than that of the interior parameters. However, we have seen that this is true for the giant planets but for the rocky planets the interior ESI is a predominant factor in the global ESI, where the real values of interior and surface ESI play a larger role than the weight exponent. The 3-D histogram in Fig. 4.2 is the result of over plotting interior and surface ESI for all the rocky exoplanets. Fig. 4.3 shows the scatter plot of interior ESI versus surface ESI. Blue dots are the giant planets, red dots are the rocky planets, and cyan circles are the solar system objects. The dashed curves are the isolines of constant global ESI. The planets above ESI = 0.8 are considered Earth-like and the planets with ESI = 0.73 are optimistically potentially habitable planets. From this, we see that there are at least 20 Earth-like planets in 3566, which may be habitable. It is also clear that interior ESI is of predominant nature. However, due to the geometrical mean nature of the global ESI formula, we need to consider all the four parameters to check the habitability of the planet. We also see from the plot that there seem to be a definite division between gaseous and rocky planets, at approximately ESI = 0.67 (interior ESI of the Moon). It is interesting to note that this division separates Moon and Io, rocky satellites, (especially Io, which is closer in bulk composition to the terrestrial planets) and, say Pluto and Europa, which are composed of water ice--rock. Previously, the ESI indexing had accepted that even planets with ESI between 0.6 and 0.8 could be potentially habitable, or at least similar to Earth. Thus, we propose to extend the optimistic limit from 0.73 to 0.67. For example, the ESI of Kepler-445 d is 0.76. It is located in the HZ and has a surface temperature of 305 K, making it suitable for life.

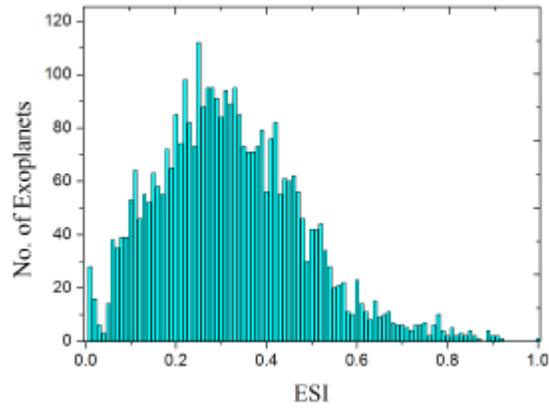

Figure 4.1: Histogram of global ESI

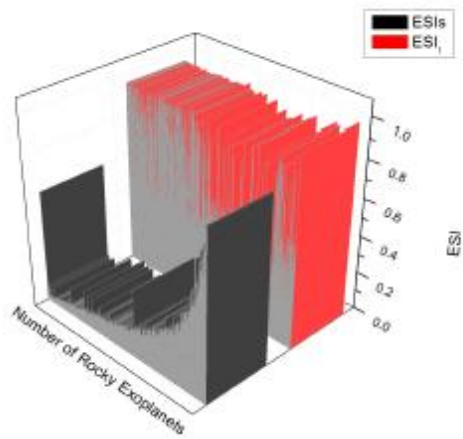

Figure 4.2: 3-D histogram of Interior and surface ESI

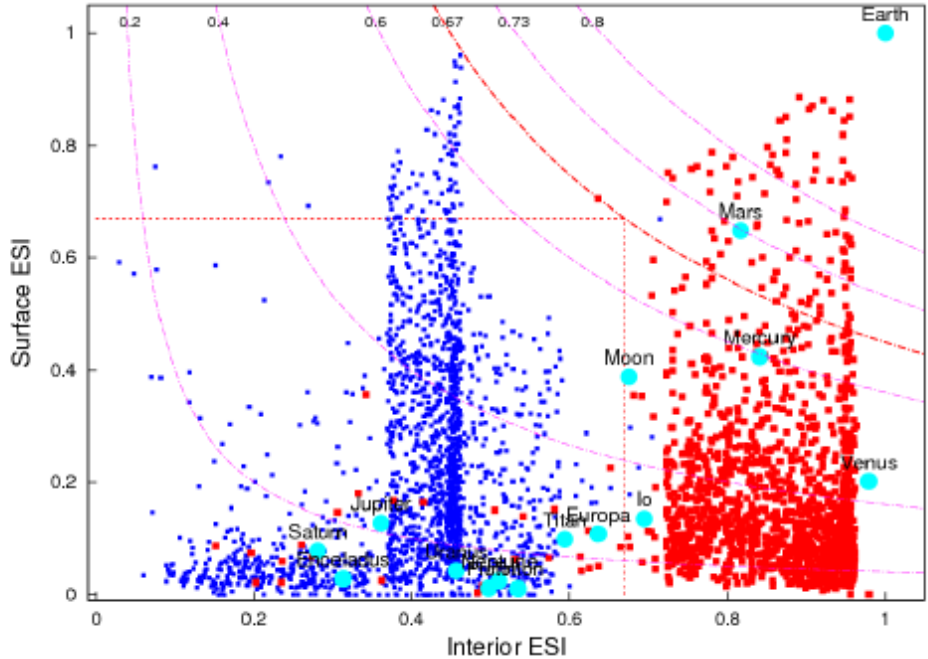

Figure 4.3: Scatter plot of Interior ESI versus surface ESI.

MSI results are shown as a histogram of the global MSI in Fig. 4.4. From the MSI results obtained, it is assumed that a planet having global MSI equal to 0.6, which is close to the value for Earth (0.68) is considered to be a Mars like planet. Fig. 4.5, shows the 3-D histogram of the interior and surface MSI. In the similar lines as that of ESI, we can see that interior MSI is more dominant factor than surface MSI for the rocky exoplanets in the global MSI.

Fig. 4.6 depicts a scatter plot of interior MSI versus surface MSI for 3566 confirmed exoplanets. The dashed curves are the isolines of constant global MSI. The planets having MSI values above 0.63 are considered Mars-like. In fig 4.7 (Left): Mass-radius diagram for exoplanets with measured masses less than 20 EU along with model curves for different mass-radius relation: black line is $R = M^{0.3}$ for $M_E > 1$; blue-dotted line is $R = M^{0.5}$ for $1 < M_E < 20$. Red crosses indicate rocky planets; blue crosses are gas giants, and cyan squares are our Solar System objects. In 2016, the data in the catalog suggested only two rocky exoplanets smaller than Earth. In the present data, there are many smaller planets. The right hand side of fig 4.7: Blow-up of the previous plot for small-size planets, in terms of the Mars units. Line of same mass-radius relation is marked on the plot, along with the isolines of constant density. Some interesting planets are marked by names.

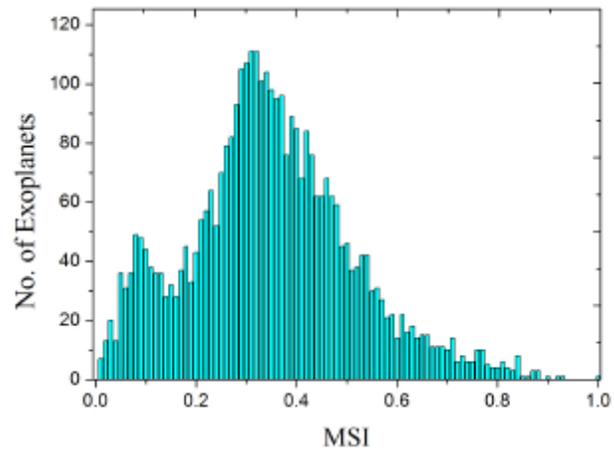

Figure 4.4: Histogram of global MSI

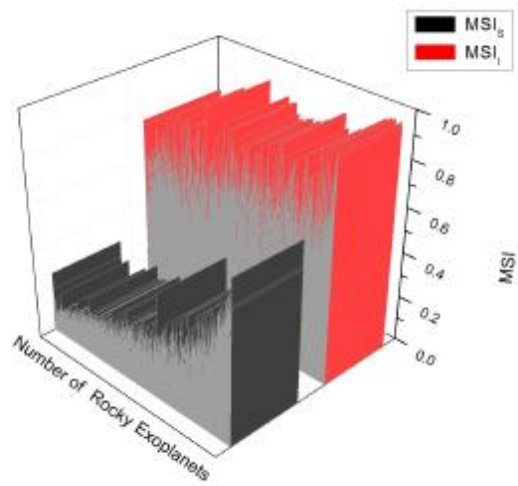

Figure 4.5: 3-D histogram of Interior and surface MSI

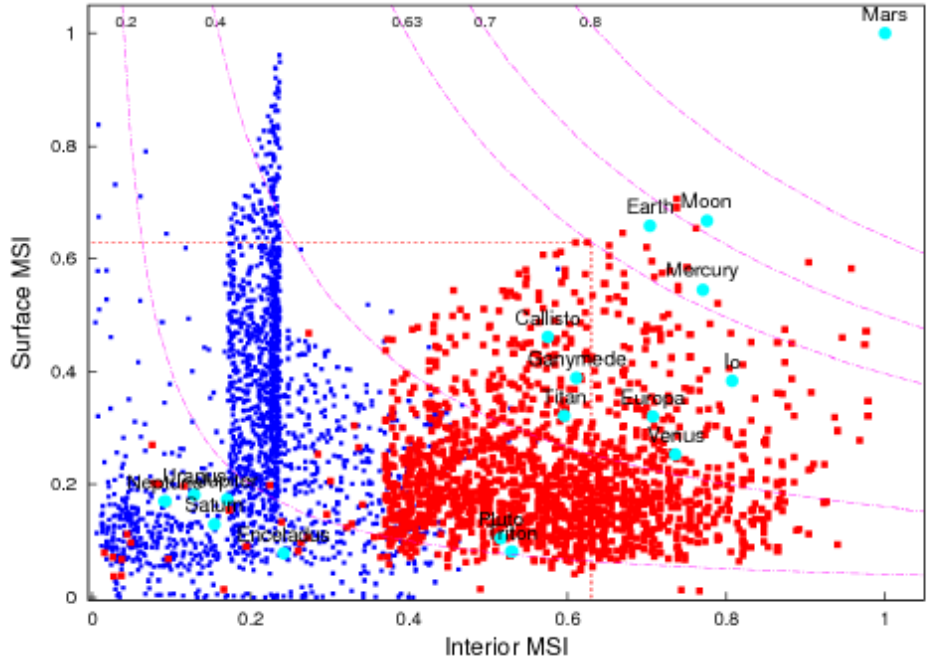

Figure 4.6: Scatter plot of interior and surface MSI

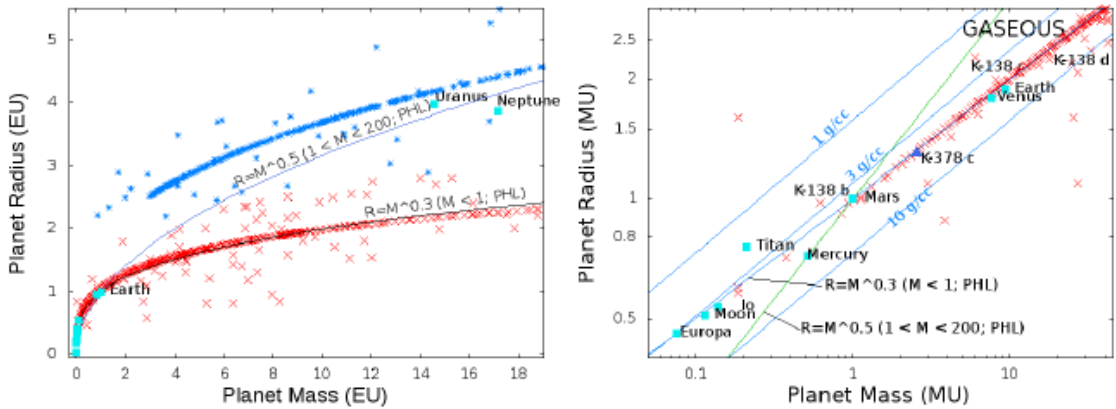

Figure 4.7 Left: Mass-radius diagram for exoplanets with measured masses less than 20 EU along with model curves for different mass-radius relation. Right: Blow-up of the previous plot for small-size planets, in terms of the Mars units.

In 2016, the data in the catalog suggested only two rocky exoplanets smaller than Earth. In the present data, there are many smaller planets.

Finally, we have used the Cobb–Douglas production function in terms of (i) Increasing parametric similarity (IPS), (ii) Decreasing parametric similarity (DPS) and ((iii) Constant parametric similarity (CPS) to understand the potential habitability of exoplanets (see Section 3.3). The sample results are presented in Table 3.6. In Figs. 4.8, 4.9 and 4.10, we have shown graphical analysis of these results. From these plots, it is clear that there is no stable variation of function. Since it does not work for varying parameters it is obvious that it also does not work for constant parametric scale (CPS). Therefore we conclude that the Cobb–Douglas production function is not suitable for potential habitability studies because it is unlike to follow the geometric mean like similarity indices does.

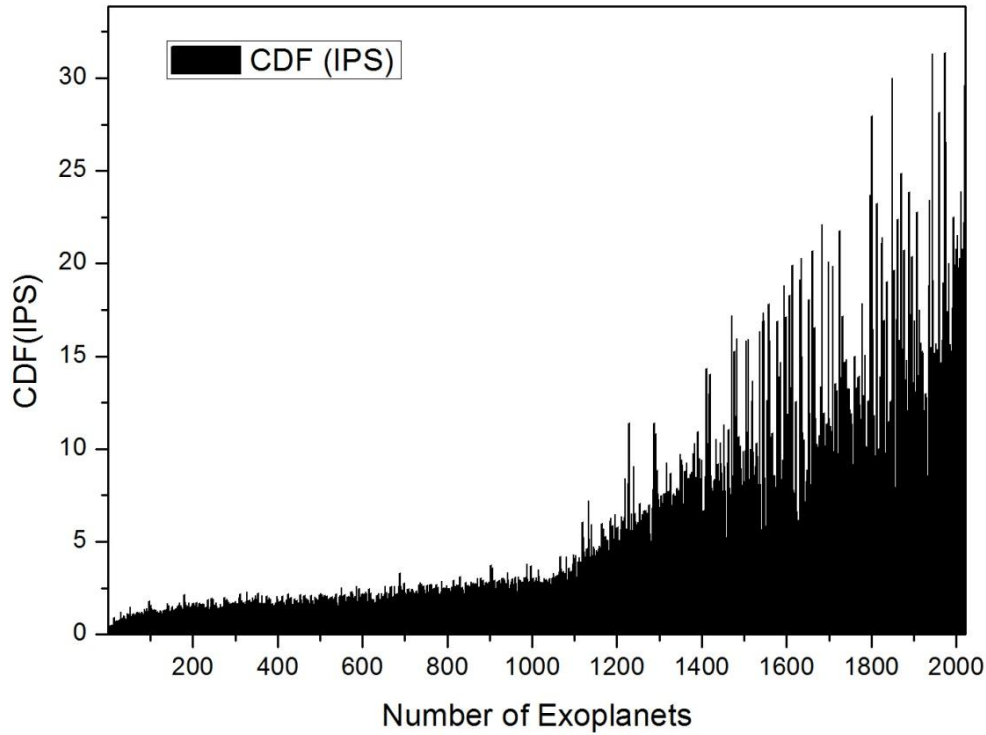

Figure 4.8: Distribution of increasing parametric scale of CDF.

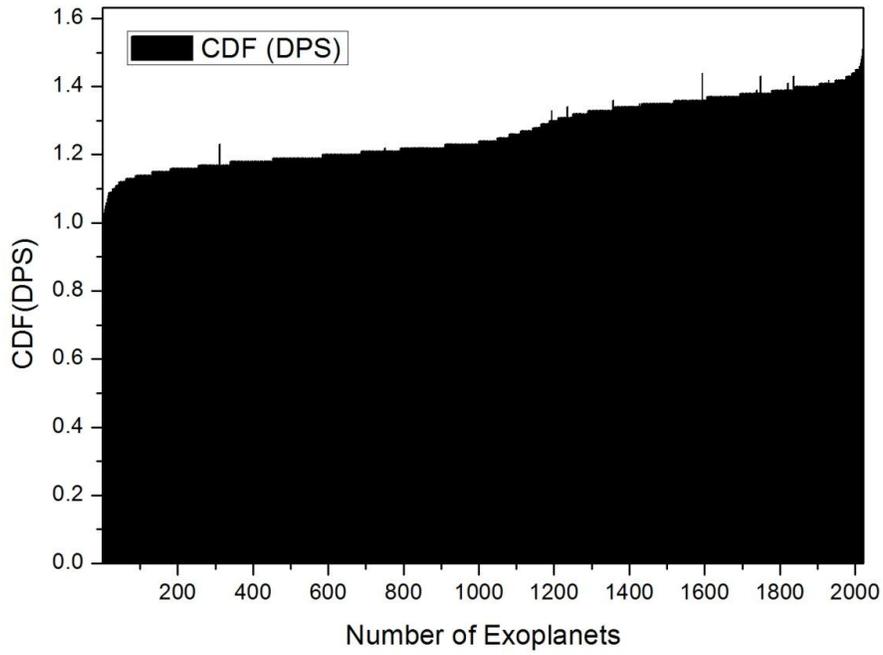

Figure 4.9: Distribution of decreasing parametric scale of CDF.

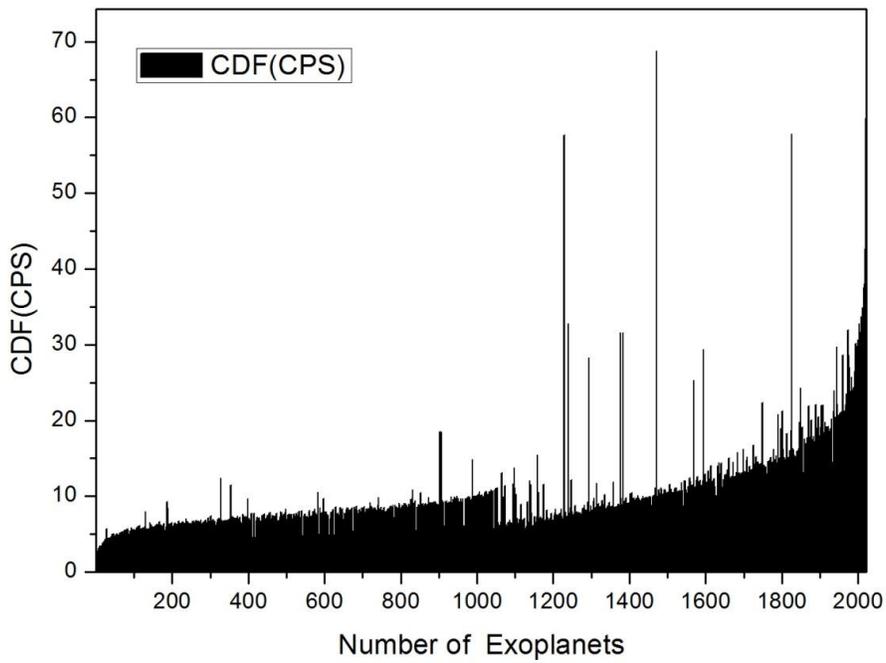

Figure 4.10: Distribution of constant parametric scale of CDF.

## 4.2 Conclusions

The search for exoplanets and habitability studies has been fascinating for mankind with the two goals, both of them have great implications to our civilization: One is to seek life elsewhere outside the Earth and another one is to seek a twin-Earth, preferably nearby. The second goal, in principle, is to have a planet habitable for our kind of life, but uninhabited, so that we can shift there in the far future. Another aspect of the second goal is that it is easier to search for the biosignatures of life as we know it on a planet which looks just like Earth.

The Earth Similarity Index (ESI), a parametric index to analyze the exoplanets' data, was introduced to access the potential habitability of all discovered to date exoplanets. In present work, we have shown how the ESI can be derived from the initial mathematical concept of similarity. Of all the four parameters entering the global ESI, only radius is a direct observable quantity while the remaining three parameters, surface temperature, escape velocity and density, are calculated. The PHL project says that surface ESI is dominating the interior ESI, because the surface temperature weight exponent value is much higher than that of the interior parameters. From this work, we found that the interior ESI is a predominant factor in the global ESI for the rocky exoplanets, where the real values of interior and surface ESI play a larger role than the weight exponents (Fig. 4.1). Due to the geometrical mean nature of global ESI formula, we need to consider all the four parameters to check the habitability of the planet. However, though evaluation of only radius and density parameters may be enough to suggest a rocky nature of an exoplanet, due to the geometrical mean nature of the ESI formulation, we need to consider the surface temperature to verify the Earth-likeness. For example, if we consider surface temperature values as 10 K, 100 K and 2500 K and keep interior ESI the same as for the Earth, the corresponding global ESI values will be 0.02, 0.40 and 0.11 respectively. Thus, the surface temperature and escape velocity do play a key role in balancing the global ESI equation. Since there is always an observational difficulty in obtaining the surface temperature value of the exoplanet directly, we introduced the calibration technique to try to mitigate this difficulty in the case of rocky planets. The planets above ESI = 0.8 may be considered Earth-like and the planets with ESI = 0.73 are optimistically potentially habitable planets, which yield at least 20 Earth-like planets from 3566 exoplanets.

From the early history of Mars, it is believed that Mars had a much wetter and warmer environment, just at the time when life on Earth is now known to have originated (this date was recently moved back to 4.1 Ga (Gigayears ago) (Bell et al. 2015). Curiosity data indicates early Martian (3.8 Ga) climate with stable water lakes on the surface for thousands to millions of years at a time (Grotzinger et al. 2015), and a recently discovered evidence of carbonate-rich (3.8 Ga) bedrock (Wray et al. 2016) suggested the habitable warm environment. It is possible that after the presumed catastrophic impact-caused loss of most of the atmosphere (Melosh and Vickery 1989; Webster et al. 2013; Webster et al. 2015; Wordsworth 2016), only the toughest life forms had survived, the ones we call here on Earth as extremophiles. They would have adapted to the currently existing conditions and just like the terrestrial extremophiles would need such conditions for the survival; for example, terrestrial methanogens have developed biological mechanism that allows them to repair DNA and protein damage to survive at temperatures from $-40\,^{\circ}\mathrm{C}$ to $145\,^{\circ}\mathrm{C}$ (Tung, Bramall and Price 2005). The usual conditions for habitability would be different for such life forms. Carbon and water have the dominant role as the backbone molecule and a solvent of biochemistry for Earth life. However, the abundance of carbon may not be a useful indication of the habitability of an exoplanet. The Earth is actually significantly depleted in carbon compared with the outer solar system. Here, on Earth, we have examples of life, both microbial and animal, that do not require large amounts of water either. For example, both bacteria and archaea are found thriving in the hot asphalt lakes (Schulze-Makuch et al. 2011b) with no oxygen and virtually no water present. They respire with the aid of metals, perhaps iron or manganese, and create their own water by breaking down hydrocarbons, just like gut bacteria that can generate most of their own water from light hydrocarbons (Kreuzer-Martin, Ehleringer and Hegg 2005). We have introduced the Mars similarity index (MSI) to study the Mars-like planets as potential planets to host extremophile life forms. In this scale, Moon has the MSI of 0.75, Earth has the MSI of 0.68, and the next closest exoplanet is Kepler-186 f (MSI = 0.69), which is listed as potentially habitable planet in the HEC. Mars-like planets can tell us about the habitability of small worlds rather than planets that are far from their star. For example, Earth at Mars distance would most probably still be habitable (Kounaves 2007). Given constant exchange of impact ejecta between the planets, it is possible that biota from the Earth reached and survived on Mars, which thus could have been `extremophile'-habitable throughout all its history.

From the work carried out, we conclude that it is necessary to arrive at the multiparameter calculator without any confinement to the number of input parameters (eg, orbital properties, temperature, escape velocity, radius, density, activation energy and so on). We would like to call this future calculator a Life Information Score (LIS), which shall be used as an overall calculator to detect life itself. The LIS is almost similar to the anthropic selection, which basically deals with the preconditions for the emergence of life and, ultimately, intelligent observers (Waltham 2011). But the expected outcome of this LIS is to accurately measure the possibility of a planet to host any form of life using only the parametric data.

Finally, our effort to use Cobb-Douglas production function to compute the planetary habitability did not yield any useful and convincing results.

Key findings of our work are summarized here. We have established a relation between surface and equilibrium temperatures using data available for the solar system objects (which have rocky surface and atmosphere) for the first time to determine mean surface temperature of exoplanets. Then we studied the ESI of all the exoplanets in our database. The study revealed that 20 Earth-like exoplanets with ESI value above 0.8 are potentially habitable planets. We have also introduced a new indexing parameter, the Mars Similarity Index (MSI), for the first time, to study the extremophile life form in Mars-like conditions. Through the graphical analysis of MSI, we have found that Moon (MSI= 0.75), Earth (MSI = 0.68) and Kepler-186f (MSI = 0.69) can be considered potentially habitable planets with MSI threshold value as 0.6.

## 4.3 Future Scope

The limitations of the present work are that we have access to only few parameters to work models and build theories based on it. In view of this, better equipped space telescopes with biosignature detectors such as PLATO, JWST are planned for future space missions (Alexander 2006). Further, understanding the astrochemistry of each planet plays an important role in improving the present understanding. This can be achieved by analyzing the chemical compositions of the planet and build a proper theory to understand there atmospheres. The ultimate goal of this type of work continues to find other life forms in this vast universe.

55. Wray J. J. et al., "Orbital evidence for more widespread carbonate-bearing rocks on Mars", *J. Geophys. Res*., vol. 121, p. 652, 2016.
44